\documentclass[11pt]{article}

\usepackage[margin=1in]{geometry}
\usepackage[utf8]{inputenc}
\usepackage{amsmath}
\usepackage{amssymb}
\usepackage{booktabs}
\usepackage{tikz}
\usetikzlibrary{arrows.meta, positioning}
\usepackage[hidelinks]{hyperref}
\usepackage{xurl}  
\usepackage{xcolor}
\usepackage{fancyhdr}

\title{Fusing Spectral Signatures and Activation Clustering for
Backdoor Detection in Healthcare Imaging Models: Method,
Implementation, and Evaluation}

\author{Suresh Tamang\\
\small Graduate Researcher, Artificial Intelligence, University of the Cumberlands\\
\small ORCID: 0009-0003-0771-6626\\
\small \texttt{suresh.tamang.dev@gmail.com}}

\date{September 6, 2026}

\begin{document}
\maketitle

\begin{abstract}
Machine learning models are increasingly deployed in healthcare imaging pipelines for diagnostic support, and training-time attacks against them are a named sector-level concern: healthcare-sector guidance identifies model poisoning and adversarial attacks as threats requiring dedicated defenses, while federal policy directs expanded AI vulnerability-detection tooling to critical infrastructure operators such as rural hospitals. Spectral signature analysis and activation clustering are two established backdoor detection methods routinely evaluated as independent baselines, but their outputs are not ordinarily combined, and reported detection performance on medical imaging benchmarks remains sparse relative to the natural-image setting. This paper contributes three things: a score-level fusion rule combining per-class spectral ranking with activation-clustering flags into a single per-sample poisoning score and a model-level agreement statistic; an open-source implementation of the resulting eight-stage pipeline; and an evaluation of that pipeline against synthetically poisoned variants of a public medical imaging benchmark and CIFAR-10 at four poisoning rates (0\%, 1\%, 5\%, 10\%) over five seeds each, measuring each detector alone against the fusion. On the medical benchmark, the fused detector reaches AUROC $\geq 0.99$ at every nonzero poisoning rate tested. On CIFAR-10, fusion does not uniformly help: at 10\% poisoning, activation clustering's true-positive rate collapses to $0.000$ and spectral AUROC independently degrades to near-chance ($0.545$), despite a $97.2\%$ attack success rate confirming the backdoor was fully installed. The fused score, a weighted combination of both signals, inherits this joint failure. Detection output is expressed in NIST AI RMF Measure-function and MITRE ATLAS terms, so findings are reported in the vocabulary security and compliance teams already use.
\end{abstract}

\section{Introduction}

AI adoption in healthcare imaging, including diagnostic support for
radiology, dermatology, and pathology, has accelerated faster than the
security tooling used to validate these systems before deployment.
Healthcare-sector guidance published in November 2025 by the Health
Sector Coordinating Council's Cybersecurity Working Group, a coalition
of more than 400 healthcare organizations, identifies model poisoning,
data corruption, and adversarial attacks as threats requiring dedicated
incident-response and secure-by-design defenses for AI-enabled medical
devices~\cite{hscc2025}.

Federal policy has moved in a related but distinct direction. Executive
Order 14409 (June 2, 2026) directs agencies to facilitate access to
cybersecurity tools and services for operators of critical
infrastructure, naming rural hospitals explicitly, and calls for determining
whether Federal grant funding is available to support advanced AI
vulnerability detection~\cite{eo14409}. The order concerns AI-enabled cyber defense
and a pre-release review process for large frontier models; it does not
itself name data poisoning or model manipulation. The threat-specific
language comes from the sector guidance, not the order, and this paper
keeps the two claims separate.

Independently, recent academic work has documented the scope of the
underlying technical vulnerability across a range of clinical AI system
designs~\cite{abtahi2026}.

This paper makes three contributions:

\begin{enumerate}
\item A score-level fusion rule combining spectral signature analysis
and activation clustering, specified with the per-class normalization
and minority-cluster guard needed to make the combination well defined
(Section~\ref{sec:method}).
\item An open-source reference implementation released under the Apache
License 2.0 (Section~\ref{sec:impl}).
\item An evaluation protocol, and its execution, measuring each detector
alone against the fusion on a medical imaging benchmark and a non-medical
benchmark at three poisoning rates
(Sections~\ref{sec:setup}--\ref{sec:results}).
\end{enumerate}

\section{Background and Related Work}

\paragraph{Data poisoning and backdoor attacks.}
Data poisoning attacks corrupt a model's training data to induce
incorrect behavior. Backdoor attacks are a stealthier subclass in which
a hidden trigger pattern causes targeted misclassification while leaving
behavior on clean inputs unaffected, making them difficult to detect
through ordinary validation-set accuracy checks. The canonical
construction injects a fixed visual trigger into a fraction of training
images and relabels them to a target class~\cite{badnets}.

\paragraph{Detection techniques and their combination.}
Tran, Li, and Madry~\cite{tran2018} show that poisoned training examples
leave a detectable signature in the spectrum of a network's learned
feature representations. Chen et al.~\cite{chen2018} cluster a network's
internal activations to separate poisoned from clean examples, on the
basis that poisoned examples activate the network via a different
internal pathway even when final-layer outputs look identical.

Stronger spectral variants exist: SPECTRE~\cite{spectre} amplifies the
spectral signature using robust covariance estimation and outperforms
the original method under several attacks. We use the 2018 formulations
as the components of the fusion because they are the versions in
widespread use as reference defenses, and note the comparison against
robust-statistics variants as future work.

Both methods are well established, and the backdoor-attack literature
routinely evaluates them together as a pair of independent baseline
defenses when measuring whether a new attack evades detection. We
therefore do not claim that jointly evaluating them is itself novel.
What is not standard, to our knowledge, is combining their outputs into
a single decision procedure: the two are typically reported as separate
detectors with separate operating points, rather than fused into one
per-sample score plus an agreement statistic. That fusion, and its
measurement against each component alone, is the methodological
contribution here.

\paragraph{Backdoors in medical imaging.}
Backdoor attacks against medical imaging models are an established
research area rather than an untouched one. Feng et al.~\cite{fiba}
introduce a frequency-domain trigger injection attack evaluated across
three medical imaging benchmarks spanning skin lesion classification,
kidney tumor segmentation, and endoscopic artifact detection, and
explicitly measure its ability to bypass existing backdoor defenses. Our
claim is accordingly narrow: the gap is not that medical-imaging
backdoors are unstudied, but that publicly available, installable
detection tooling reporting results in framework-referenced terms for
healthcare security teams is scarce, and that the fused decision
procedure evaluated here has not been characterized on medical imaging
data.

\paragraph{Policy and governance context.}
The NIST AI Risk Management Framework organizes AI risk management into
four functions, Govern, Map, Measure, and Manage, of which Measure
(technical testing and quantification) is the function this work
serves~\cite{nistairmf}. Governance frameworks such as ISO/IEC
42001~\cite{iso42001}
provide organizational policy structure but do not themselves supply
technical validation evidence. MITRE ATLAS catalogs adversary tactics
and techniques against AI systems in a structure analogous to MITRE
ATT\&CK~\cite{atlas}. The attack simulated and detected here maps to
three ATLAS techniques, reported together in Section~\ref{sec:results}:
\textbf{AML.T0020} (poisoning of training data, the technique stage 02
simulates directly)~\cite{atlas-t0020}, \textbf{AML.T0059} (erosion of
dataset integrity, of which a crafted backdoor is a targeted special
case)~\cite{atlas-t0059}, and \textbf{AML.T0018}, \emph{Backdoor ML
Model} (the persistent, hidden change in the trained model's behavior
that a successful poisoning attack produces)~\cite{atlas-t0018}.

\section{Detection Method}
\label{sec:method}

Let $A_c \in \mathbb{R}^{n_c \times d}$ be the matrix of
intermediate-layer activations for the $n_c$ training samples labeled
class $c$.

\paragraph{Spectral signature score.}
Center $A_c$ to obtain $\tilde{A}_c = A_c - \bar{a}_c$, and let $v_c$ be
the top right singular vector of $\tilde{A}_c$. Sample $i$ receives
$s_i = \langle \tilde{a}_i, v_c \rangle^2$. Raw scores are not
comparable across classes, so each is replaced by its within-class
percentile rank $\hat{s}_i \in [0,1]$.

\paragraph{Activation clustering flag.}
Project $A_c$ to $10$ dimensions by PCA (seeded) and partition by
$k$-means with $k = 2$. Let $m_c$ be the smaller cluster. Set the flag $f_i = 1$ for
$i \in m_c$ if and only if $|m_c| / n_c \le \tau$, and $f_i = 0$
otherwise. The guard $\tau$ (default $0.35$) is required because
$k$-means with $k=2$ always returns a partition, including for a class
containing no poisoned samples; without it the method reports a spurious
minority cluster in every clean class.

\paragraph{Fusion and decision rule.}
The per-sample fused score is
$g_i = \tfrac{1}{2}\hat{s}_i + \tfrac{1}{2}f_i$, giving a continuous
ranking. Two boolean outputs are also reported: the clustering flag
$f_i$ alone, and the agreement indicator
$f_i \wedge [\hat{s}_i \ge \theta]$, with
$\theta = 0.9$, i.e.\ a sample counts as spectrally flagged when its
within-class percentile rank falls in the top 10\% of its class. The
cutoff affects only the agreement indicator; the continuous fused score
$g_i$ is unaffected by it.
The model-level risk statistic is the fraction of samples satisfying the
agreement indicator. Requiring agreement necessarily reduces the number of flagged samples
relative to either detector alone, so recall cannot increase. Whether
precision improves is an empirical question, not a mathematical
guarantee, and is measured in Section~\ref{sec:results} rather than
assumed. Likewise, the guard $\tau$ suppresses the degenerate
all-classes-flagged behavior but does not eliminate false positives on
clean data, which is why $0\%$-poisoning controls are included in the
evaluation. Reporting both operating points rather than a single number
is deliberate: an assurance report that collapses them hides this
trade-off.

\section{Implementation}
\label{sec:impl}

\begin{figure}[t]
\centering
\begin{tikzpicture}[
  node distance=3.2mm,
  stage/.style={draw, rounded corners=2pt, align=center,
                minimum height=8mm, text width=30mm, font=\scriptsize},
  arr/.style={-{Stealth[length=2mm]}, thick}
]
\node[stage] (s1) {01 Load benchmarks};
\node[stage, right=of s1] (s2) {02 Inject synthetic\\backdoor triggers};
\node[stage, right=of s2] (s3) {03 Train classifier};
\node[stage, right=of s3] (s4) {04 Extract activations};

\node[stage, below=9mm of s1] (s5) {05 Spectral analysis\\+ activation clustering};
\node[stage, right=of s5] (s6) {06 Fuse scores};
\node[stage, right=of s6] (s7) {07 Evaluate\\TPR / FPR / AUROC};
\node[stage, right=of s7] (s8) {08 Map to ATLAS\\+ NIST AI RMF};

\draw[arr] (s1) -- (s2);
\draw[arr] (s2) -- (s3);
\draw[arr] (s3) -- (s4);
\draw[arr] (s4.south) -- ++(0,-3mm) -| (s5.north);
\draw[arr] (s5) -- (s6);
\draw[arr] (s6) -- (s7);
\draw[arr] (s7) -- (s8);

\node[align=center, font=\tiny, below=7mm of s7] (gt)
  {ground-truth poisoning mask\\from stage 02 (evaluation only)};
\draw[arr, dashed] (gt) -- (s7);
\end{tikzpicture}
\caption{The eight-stage pipeline. Stage 02 produces the ground-truth
poisoning mask, which is withheld from stages 05 and 06 and consumed
only at stage 07.}
\label{fig:pipeline}
\end{figure}
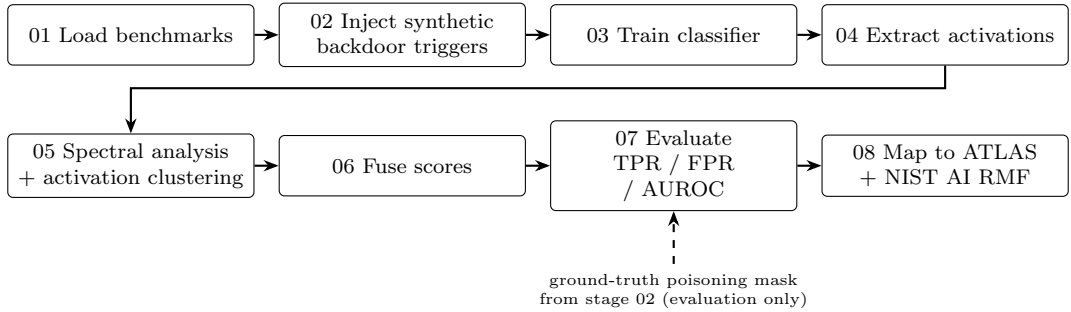

The method is implemented as \texttt{aegis-scan}, an open-source,
pip-installable command-line tool released under the Apache License 2.0
at \url{https://github.com/tsurace37/aegis-scan}, in Python 3.11 with
PyTorch. Intermediate activations are captured through forward hooks
that are detached after use, so extraction does not alter the model's
inference behavior. The pipeline is organized into the eight stages in
Figure~\ref{fig:pipeline}.

The detection and fusion stages never observe the ground-truth poisoning
mask produced at injection time; it is consumed only by the evaluation
stage. This separation is enforced at the interface level so that
reported performance cannot be inflated by label leakage.

\section{Experimental Setup}
\label{sec:setup}

\paragraph{Datasets.}
Two public benchmarks are used. The medical benchmark is
PneumoniaMNIST from the MedMNIST v2 collection~\cite{medmnist},
a binary chest radiograph classification task at $28 \times 28$
resolution. The non-medical benchmark is CIFAR-10~\cite{cifar10}, 
matching the setting in which the two detection methods were originally
evaluated. The MedMNIST collection is distributed under CC BY 4.0 (DermaMNIST,
which is not used here, is CC BY-NC 4.0). CIFAR-10 is distributed by the
University of Toronto for research and educational use; no separate
data-use agreement is required.

\paragraph{Threat model.}
The adversary can modify a bounded fraction of training samples,
inserting a trigger and relabeling those samples to a fixed target
class. The adversary cannot modify the architecture, training procedure,
or loss. The defender has white-box access to the trained model and to
the full (possibly poisoned) training set, but no knowledge of the
trigger, the target class, or the poisoning rate.

\paragraph{Attack construction.}
Poisoning follows the BadNets construction~\cite{badnets}: a
$3\times3$ solid square patch at maximum intensity ($1.0$ in the
normalized $[0,1]$ range) is written into the bottom-right corner of the
image, inset by a $2$-pixel margin, and the label is set to target class
$t = 0$ (the implementation default) for both datasets. Only samples not already labeled $t$ are poisoned, so that the
label flip carries signal. Poisoning rates of 1\%, 5\%, and 10\% of the
training set are evaluated.

\paragraph{Model and training.}
The classifier is a compact, hand-written residual
network~\cite{resnet}, not a standard ImageNet-scale architecture: a
$3\times3$ stem convolution (producing $32$ channels) feeds three
residual blocks at strides $1$, $2$, $2$ respectively, with channel width
held at $32$ through the first (stride-$1$) block and doubled at each of
the two downsampling blocks ($32 \to 64 \to 128$), followed by global
average pooling and a linear classifier head. This is sized deliberately for
$28$--$32$ pixel inputs, rather than a standard torchvision ResNet
sized for $224\times224$ ImageNet input: this architecture preserves
more spatial resolution at these smaller input sizes and requires less
computation than repurposing an ImageNet-scale stem, since a
$28\times28$ input does not need the aggressive early downsampling a
$224\times224$-input stem is designed for. The network is
trained with Adam and cross-entropy loss for $10$ epochs at learning
rate $10^{-3}$, batch size $64$, with no weight decay and no data
augmentation. Activations for detection are taken from the third
residual block (128 channels), resolved by module name via
\texttt{model.get\_submodule("layer3")} and flattened to one row per
sample. These are the implementation's code defaults and were used
unmodified for every run reported here. MedMNIST publishes train, validation, and
test partitions; only the train and test partitions are used, since the
pipeline performs no validation-based model selection or early stopping
--- training runs for a fixed epoch count regardless of held-out
performance, and the implementation as of the commit cited below does
not consume a validation split at all. Detection operates on the
training split only; the test split is reserved for the clean-accuracy
and attack-success-rate measurements in Section~\ref{sec:results}.

\paragraph{Repetition and uncertainty.}
Each configuration is run with $5$ independent random seeds
($0,1,2,3,4$); tables report mean and standard deviation. Results were
produced with \texttt{aegis-scan} at commit
\texttt{6bf3af4c46ede67d48f3a3bbb2cec801a348fccb}, so that the exact
code state behind every number is identifiable. That commit includes two
fixes made in the course of running the full grid: the spectral-signature
stage's singular value decomposition was switched from NumPy's default
\texttt{gesdd} driver
to SciPy's \texttt{gesvd}, because \texttt{gesdd} reliably raised an
out-of-memory error on the CIFAR-10 activation matrices (approximately
$5000 \times 8192$ per class) on the Windows machine used to run these
experiments, even with ample RAM free; and a crash in the results-table
generation script when a metric is legitimately undefined (see
Section~\ref{sec:results}) was fixed so that such cases report as
explicitly undefined rather than halting the script.

\section{Results}
\label{sec:results}

Tables~\ref{tab:results-accuracy} and~\ref{tab:results-detection}
report the full grid: two datasets, four
poisoning rates (a $0\%$ false-positive control plus $1\%$, $5\%$, and
$10\%$ attacks) --- eight dataset/rate configurations, each run with
five seeds, forty runs in total.

\begin{table}[t]
\centering
\small
\begin{tabular}{llcc}
\toprule
\textbf{Dataset} & \textbf{Rate} & \textbf{ASR} & \textbf{Clean acc.} \\
\midrule
PneumoniaMNIST & 0\% (control) & 0.029 $\pm$ 0.033 & 0.866 $\pm$ 0.030 \\
PneumoniaMNIST & 1\%  & 0.989 $\pm$ 0.017 & 0.869 $\pm$ 0.019 \\
PneumoniaMNIST & 5\%  & 0.999 $\pm$ 0.001 & 0.862 $\pm$ 0.027 \\
PneumoniaMNIST & 10\% & 1.000 $\pm$ 0.000 & 0.756 $\pm$ 0.162 \\
\addlinespace
CIFAR-10 & 0\% (control) & 0.039 $\pm$ 0.026 & 0.779 $\pm$ 0.008 \\
CIFAR-10 & 1\%  & 0.917 $\pm$ 0.029 & 0.768 $\pm$ 0.013 \\
CIFAR-10 & 5\%  & 0.965 $\pm$ 0.008 & 0.762 $\pm$ 0.020 \\
CIFAR-10 & 10\% & 0.972 $\pm$ 0.004 & 0.762 $\pm$ 0.018 \\
\bottomrule
\end{tabular}
\caption{Attack success and clean-test accuracy, mean $\pm$ sd over 5
seeds. \textbf{ASR} is attack success rate on triggered, held-out,
non-target-class test inputs. \textbf{Clean acc.} is each row's own
trained model evaluated on the ordinary (untriggered) test set --- the
same model used for that row's ASR figure, not a separately trained
comparison model; the $0\%$ rows, trained with no poisoning at all,
supply the unpoisoned baseline that the attacked rows are compared
against. PneumoniaMNIST's $10\%$ row shows accuracy does not hold at
the unpoisoned baseline uniformly across rates
(Section~\ref{sec:results} discusses this directly).}
\label{tab:results-accuracy}
\end{table}

\begin{table}[t]
\centering
\small
\begin{tabular}{llcccc}
\toprule
\textbf{Dataset} & \textbf{Rate} & \textbf{Spectral} & \textbf{Clustering} &
\textbf{Fused} & \textbf{Agreement} \\
 & & AUROC & TPR & AUROC & Precision \\
\midrule
PneumoniaMNIST & 0\% (control) & n/a & n/a & n/a & n/a \\
PneumoniaMNIST & 1\%  & 0.986 $\pm$ 0.000 & 0.996 $\pm$ 0.010 & 1.000 $\pm$ 0.000 & 1.000 $\pm$ 0.000 \\
PneumoniaMNIST & 5\%  & 0.940 $\pm$ 0.001 & 0.998 $\pm$ 0.002 & 0.999 $\pm$ 0.001 & 1.000 $\pm$ 0.000 \\
PneumoniaMNIST & 10\% & 0.894 $\pm$ 0.004 & 0.995 $\pm$ 0.004 & 0.996 $\pm$ 0.004 & 1.000 $\pm$ 0.000 \\
\addlinespace
CIFAR-10 & 0\% (control) & n/a & n/a & n/a & 0.000 (n=2/5) \\
CIFAR-10 & 1\%  & 0.934 $\pm$ 0.004 & 0.916 $\pm$ 0.012 & 0.968 $\pm$ 0.005 & 1.000 $\pm$ 0.000 \\
CIFAR-10 & 5\%  & 0.811 $\pm$ 0.005 & 0.960 $\pm$ 0.005 & 0.963 $\pm$ 0.005 & 1.000 $\pm$ 0.000 \\
CIFAR-10 & 10\% & 0.545 $\pm$ 0.007 & 0.000 $\pm$ 0.000 & 0.545 $\pm$ 0.007 & n/a \\
\bottomrule
\end{tabular}
\caption{Detection performance, mean $\pm$ sd over 5 seeds, on the same
40 runs as Table~\ref{tab:results-accuracy}. \textbf{Clustering} is
reported as TPR rather than AUROC, since it is a binary flag rather than
a continuous score. The $0\%$ rows are false-positive controls, and each
reported statistic's definedness there depends on its own denominator,
not on a single shared condition: \textbf{TPR} ($\text{TP}/(\text{TP}+\text{FN})$)
is undefined at $0\%$ poisoning because $\text{TP}+\text{FN}=0$ always ---
no actual positives exist to recall. \textbf{AUROC} is undefined for the
same underlying reason (ground truth contains only the negative class).
\textbf{Precision} ($\text{TP}/(\text{TP}+\text{FP})$) is different: $\text{TP}=0$
is guaranteed at $0\%$, but precision is undefined only when
$\text{FP}=0$ too; whenever anything at all is flagged on clean data,
precision is a defined $0$, not undefined. CIFAR-10's agreement
precision at $0\%$ demonstrates this directly: it is $\mathbf{0.000}$
(defined) on 2 of 5 seeds, where something was flagged and flagged
incorrectly, and undefined on the other 3, where nothing was flagged at
all. \textbf{FPR} ($\text{FP}/(\text{FP}+\text{TN})$) is undefined only
when $\text{FP}+\text{TN}=0$, which cannot happen on a nonempty clean
dataset, since every sample is a true negative or a false positive there
--- it is always defined at $0\%$ poisoning. PneumoniaMNIST's clustering
FPR at $0\%$ is confirmed at exactly $0.000$ across all 5 seeds
($0/4{,}708$ in every case). CIFAR-10's is not uniformly zero: $0.000$
on 3 of 5 seeds, but $3.46\%$ and $3.00\%$ on the other two ($1{,}731$
and $1{,}499$ of $50{,}000$ clean images respectively) --- exactly the 2
seeds where agreement precision above is a defined $0.000$ rather than
undefined, confirming the logical link between the two columns. This
seed-dependent clean-data false-positive pattern on CIFAR-10, absent on
PneumoniaMNIST, is discussed in Section~\ref{sec:results} and is not yet
explained.}
\label{tab:results-detection}
\end{table}

\paragraph{The backdoor was genuinely installed.} Attack success rate
climbs sharply with poisoning rate on both datasets and is not close to
the $0\%$-control baseline even at $1\%$ poisoning (PneumoniaMNIST
$98.9\%$, CIFAR-10 $91.7\%$), reaching effectively total success by
$10\%$ (PneumoniaMNIST $100.0\%$, CIFAR-10 $97.2\%$). This matters for
the paper's own claim: a detection result is only informative if there
was something real to detect, and Table~\ref{tab:results-accuracy}
establishes that there was.

\paragraph{Fusion improves on spectral alone on the medical benchmark.}
On PneumoniaMNIST, fused AUROC is $\geq 0.996$ at every attacked
poisoning rate, exceeding the spectral detector alone (which ranges from
$0.894$ to $0.986$) at every rate --- a valid comparison, since both are
reported on the same continuous-score AUROC metric. Clustering alone is
already strong on this benchmark (TPR $\geq 0.995$ throughout), but we
do not claim the continuous fused score $g_i$ improves on clustering:
AUROC and TPR are different metrics computed from different output
types (continuous score vs.\ boolean flag), and comparing them directly
is not meaningful. A separate output, the \emph{agreement indicator}
$f_i \wedge [\hat{s}_i \ge \theta]$ (Section~\ref{sec:method}), is a
boolean decision rule distinct from $g_i$ and is where a same-metric,
boolean-vs-boolean comparison against clustering alone is possible.
Figure~\ref{fig:report} shows that comparison does not favor agreement:
clustering alone reaches $100\%$ recall at $100\%$ precision on that
run, while the agreement indicator drops recall to $61.7\%$ at the same
precision, a straightforward recall cost with no precision gain in that
instance. This is a limitation of the agreement rule specifically, not
evidence about the continuous fused score $g_i$, which is not evaluated
against clustering on a shared metric anywhere in this paper. The
correctly supported claim is: fusion ($g_i$) helps relative to spectral
alone; agreement's relationship to clustering alone is unfavorable in
the one instance measured here; and $g_i$'s relationship to clustering
alone is not established either way.

\paragraph{Both detectors independently fail at CIFAR-10 10\%, and
this is the paper's most important finding.} At this configuration,
activation clustering's true-positive rate is $0.000 \pm 0.000$ across
all five seeds, and spectral AUROC is $0.545 \pm 0.007$, barely above
the $0.500$ chance level. This is not a case of fusion damaging a
working detector: the fused score is $g_i = 0.5\hat{s}_i + 0.5 f_i$, so
if the clustering flag $f_i$ is zero for every sample, $g_i$ reduces
exactly to a positive scalar multiple of the spectral score $\hat{s}_i$,
and AUROC is invariant to positive rescaling --- fused AUROC would equal
spectral AUROC exactly, not approximately, which is what the measured
values ($0.545$ and $0.545$) are consistent with. We confirmed this
directly rather than leaving it as a plausible reading of TPR$=0$: a
diagnostic script cross-referencing each run's saved ground-truth mask
against its saved per-sample clustering flags shows the flag is exactly
zero for every one of the $50{,}000$ training samples, in every class,
in all five seeds at this configuration --- not merely zero recall on
poisoned samples, but zero flags anywhere. The fused-AUROC-equals-
spectral-AUROC claim is accordingly established, not merely consistent
with the data.

Separately, and regardless of that confirmation: this is not a case
where the attack failed and detection correctly found nothing. ASR at
this configuration is $97.2\% \pm 0.4\%$, confirming the backdoor was
installed essentially perfectly. The detector's blind spot and the
attack's success are simultaneous, independent facts, and a report that
surfaced only the fused AUROC number at this configuration would
materially misrepresent the system's actual security value here.

\paragraph{A guard-threshold explanation consistent with the confirmed
outputs.} The minority-cluster guard $\tau = 0.35$ described in
Section~\ref{sec:method} exists to suppress $k$-means's tendency to
always produce a two-way split even in a fully clean class, by refusing
to flag a cluster unless it is genuinely a minority of its class. The
relevant quantity is not dataset size or absolute sample counts, but the
poisoned fraction \emph{within the target class} after poisoning:
writing $p$ for the poisoning rate as a fraction of the full dataset and
$q$ for the target class's original (pre-poisoning) fraction of the
dataset, the target class ends up $q + p$ of the dataset, of which $p$
is poisoned, so the within-class contaminated fraction is $p/(q+p)$.
CIFAR-10's ten classes are exactly balanced ($q = 0.10$); at $p = 0.10$
poisoning this gives $p/(q+p) = 0.10/0.20 = 50\%$ contamination of the
target class --- comfortably \emph{above} $\tau$. \emph{If} $k$-means's
own two-way split in that class ends up similarly sized to the true
poisoned/clean split, the guard would predict suppression of the
minority cluster rather than a flag; whether $k$-means's split actually
tracks contamination this closely is the open question this section
returns to below.

What we confirmed directly, via a diagnostic script cross-referencing
each run's saved ground-truth mask against its saved per-sample
clustering flags, is the measured contamination fractions and the
measured flag counts, not the internal cluster assignment that produced
them. For CIFAR-10 at $10\%$ poisoning, the target class contains
exactly $10{,}000$ post-poisoning samples ($50.0\%$ contamination as
predicted), and the measured flag count --- in that class and every
other class --- is exactly zero across all five seeds. For
PneumoniaMNIST, the target class contained $1{,}214$ samples before
poisoning ($q = 25.8\%$) and $471$ newly poisoned samples, giving a
measured contamination of $28.0\%$, matching the formula's prediction
($27.9\%$) to within rounding and comfortably under $\tau$; here
clustering flags $470/471$, $468/471$, $469/471$, $471/471$, and
$466/471$ poisoned samples across the five seeds, with zero false
positives in every one, so the flagged set and the truly poisoned set
coincide almost exactly. This near-total overlap on PneumoniaMNIST is
evidence, though not direct inspection, that the two $k$-means clusters
in that class do track poisoning status when the guard permits a flag
at all. For CIFAR-10 at $10\%$, no equivalent evidence is available:
the guard suppresses the flag before it can be compared against
anything, so while the guard's known threshold rule offers a plausible
explanation for why zero samples are flagged at $50\%$ measured
contamination --- \emph{if} $k$-means's own cluster split ends up
similarly sized to the poisoned/clean split, which we have not verified
--- we have not verified whether the underlying, suppressed $k$-means
partition would have separated poisoned from clean samples had the
guard allowed it through. We accordingly describe the guard-threshold account as
consistent with the confirmed contamination and flag-count measurements
--- established facts about the input and output of the detector ---
rather than as a confirmed account of what $k$-means's internal
partition actually contained. This explanation addresses clustering's
failure only; it says nothing about why spectral signature analysis,
which does not use $\tau$ or any per-class minority concept,
independently degrades to near-chance at the same CIFAR-10 configuration
(Section~\ref{sec:limitations} lists this as unexplained).

\paragraph{An unexplained, seed-dependent false-positive burden on
CIFAR-10 clean data.} The same diagnostic pass surfaced a second finding
we did not anticipate. At $0\%$ poisoning, PneumoniaMNIST's clustering
false-positive rate is exactly $0.000$ on all five seeds, with no
exceptions. CIFAR-10's is not: two of five seeds (seeds $0$ and $1$)
show substantial false-positive rates of $3.46\%$ and $3.00\%$
($1{,}731$ and $1{,}499$ of $50{,}000$ clean training images flagged),
while the remaining three seeds show exactly $0.000$, matching
PneumoniaMNIST's pattern. This is not a small-count rounding artifact:
over a thousand images are involved in each affected seed. We do not
have an explanation for why two of five otherwise-identical training
runs on entirely clean data produce this behavior and three do not; it
is plausible that some seeds' trained models develop activation-space
substructure unrelated to poisoning (near-duplicate images, visual
subclasses within a coarse label, or similar) that a per-class $k$-means
split occasionally partitions along, clearing $\tau$ by chance rather
than by design. This is a real operational concern independent of the
poisoning-detection question this paper otherwise addresses: a security
team using this tool on CIFAR-10-like data could see a $3\%$
false-alarm rate on a clean dataset purely as a function of which
training seed was used, with no poisoning present at all. We report
this rather than average over it, and list understanding its cause as
future work (Section~\ref{sec:limitations}).

\paragraph{One low-accuracy seed, cause unknown.} PneumoniaMNIST clean
accuracy at $10\%$ poisoning is $0.756 \pm 0.162$ against a
$0.866 \pm 0.030$ unpoisoned baseline (Table~\ref{tab:results-accuracy},
$0\%$ row). The five individual per-seed values, recovered directly from
each run's saved output, are $0.808$, $0.468$, $0.832$, $0.857$, and
$0.817$: one seed (seed $1$) reached $46.8\%$, which is below the
$64.2\%$ trivial baseline of always predicting the majority class
(\texttt{pneumonia}); the other four averaged $82.8\%$, against a
five-seed clean baseline of $86.6\%$. We report these figures as
measured and do not offer an explanation for seed $1$'s low accuracy:
falling below the trivial baseline is consistent with a training run
that failed to converge, but we have not inspected training curves or
any other evidence that would distinguish that from another cause, and
do not claim to know why. Nor do we claim the other four seeds establish
that the backdoor leaves accuracy fully intact at this configuration:
their $82.8\%$ average sits below the $86.6\%$ unpoisoned baseline as
well, a smaller gap that this study does not further characterize. All
five seeds are retained in Table~\ref{tab:results-accuracy}'s reported
mean and standard deviation; identifying seed $1$ here explains the
table's unusually large variance without treating any seed as an outlier
to be excluded.

\paragraph{Framework-referenced reporting.}
Stage 08 emits a self-contained Markdown assurance report. It performs
no analysis of its own: every number in it traces back to a stage 07
measurement against ground truth, so the stage is a translation layer
rather than an additional source of evidence. The report has a fixed
structure --- a one-sentence plain-language finding, the detection
metrics, and the two framework mappings in Table~\ref{tab:frameworks}
--- with only the numeric fields varying by run. The headline sentence
selects the strongest available evidence, preferring the fused
continuous score and falling back through the agreement indicator,
clustering, and spectral scores depending on which detector outputs were
supplied.

\begin{table}[t]
\centering
\small
\begin{tabular}{p{0.22\textwidth}p{0.7\textwidth}}
\toprule
\textbf{Framework reference} & \textbf{Mapping asserted by the report} \\
\midrule
MITRE ATLAS \textbf{AML.T0020} & The technique the tool is built around;
stage 02 simulates it directly by modifying a subset of training images
and their labels, and stages 05--07 are the detection layer for it. \\
\addlinespace
MITRE ATLAS \textbf{AML.T0059} & A poisoned dataset is a targeted
special case: the altered subset is not random corruption but is crafted
to survive training and re-emerge as a backdoor, which is why basic
data-quality checks do not catch it. \\
\addlinespace
MITRE ATLAS \textbf{AML.T0018} (\emph{Backdoor ML Model}) & The end
state of a successful attack, a persistent hidden change in model
behavior. The tool does not inspect model weights directly as the
sub-technique AML.T0018.000 (\emph{Poison ML Model}) describes; it
infers the same outcome indirectly from how poisoned data reshapes a
layer's activations. \\
\addlinespace
NIST AI RMF \textbf{MEASURE 2.7} & ``AI system security and resilience
--- as identified in the MAP function --- are evaluated and
documented.'' The report is that evaluation, scoped to one named,
testable risk and quantified against known ground truth, rather than a
policy statement that testing occurred. \\
\bottomrule
\end{tabular}
\caption{The framework mappings stage 08 asserts. These are fixed
properties of the tool, not per-run results. Deliberately narrow: a
loose mapping that claims broad framework coverage is the failure mode
this reporting layer exists to avoid.}
\label{tab:frameworks}
\end{table}

Figure~\ref{fig:report} reproduces one such report verbatim (PneumoniaMNIST,
5\% poisoning, seed 0), rather than only describing the schema above.

\begin{figure}[t]
\small
\begin{verbatim}
# aegis-scan Assurance Report

Dataset: healthcare
Generated: 2026-09-07T12:18:13.107608+00:00
Poison rate (ground truth): 4.992%

## Finding

The fused detector (spectral signature analysis + activation clustering
combined) achieved AUROC=1.000 (excellent) and caught 100.0% of truly
poisoned samples when flagging exactly as many samples as were
actually poisoned.

## Detection metrics

- Activation clustering: TPR=100.0%, FPR=0.0%, precision=100.0%
  (tp=235, fp=0, fn=0, tn=4473)
- Both detectors agree: TPR=61.7%, FPR=0.0%, precision=100.0%
  (tp=145, fp=0, fn=90, tn=4473)
- Spectral signature analysis: AUROC=0.941 (excellent),
  average precision=0.312, top-k recall=31.1%
- Fused score (spectral + clustering combined): AUROC=1.000 (excellent),
  average precision=1.000, top-k recall=100.0%

## MITRE ATLAS mapping

- [AML.T0020] Poison Training Data
- [AML.T0059] Erode Dataset Integrity
- [AML.T0018] Manipulate AI Model

## NIST AI RMF mapping

MEASURE 2.7: "AI system security and resilience -- as identified in
the MAP function -- are evaluated and documented."
\end{verbatim}
\caption{The literal output of stage 08 for one configuration
(PneumoniaMNIST, 5\% poisoning, seed 0). ATLAS and NIST mapping bodies
truncated here for space; the full text matches Table~\ref{tab:frameworks}.}
\label{fig:report}
\end{figure}

Note the divergence between the two boolean detector outputs on this
single run: clustering alone reaches $100\%$ recall with zero false
positives, while requiring both detectors to agree drops recall to
$61.7\%$ at the same zero false-positive rate. This is the precision/recall
trade-off described in Section~\ref{sec:method}, visible in a real
report rather than only asserted in the abstract. The report's top-$k$
recall figures (spectral $31.1\%$, fused $100.0\%$ on this run) use an
oracle evaluation budget: recall when flagging exactly as many samples
as are actually poisoned. This is a standard way to evaluate ranking
quality independent of a threshold choice, but it assumes knowledge a
deployed detector does not have --- the true poisoning count --- and
should not be read as an operating point a defender could reproduce in
practice. The TPR/FPR/precision figures elsewhere in the report, which
use the fixed decision rules from Section~\ref{sec:method}, are the
operational metrics; top-$k$ recall is an evaluation diagnostic only.

AML.T0018 is \emph{Backdoor ML Model}~\cite{atlas-t0018}, confirmed
against current MITRE ATLAS documentation and third-party sources citing
it as such as recently as May 2026; this is the name used
throughout this paper. The literal report text reproduced in Figure~2
predates this check and still shows the implementation's internal,
non-standard label (``Manipulate AI Model''); the underlying technique
ID (AML.T0018) is correct and unaffected, but the display string in
\texttt{aegis-scan}'s stage 08 reporting code should be corrected to
match ATLAS's actual terminology, which we list as an open
implementation fix in Section~\ref{sec:limitations} rather than silently
editing the reproduced report text after the fact.

\section{Discussion and Limitations}\label{sec:limitations}

Evaluation on synthetically poisoned public benchmarks demonstrates
detection capability under controlled conditions. It does not establish
performance against adversarially optimized poisoning designed to evade
known detection signatures; frequency-domain attacks such
as~\cite{fiba} are explicitly constructed to survive existing defenses,
and are not evaluated here. This constraint is common in the
backdoor-detection literature given the difficulty of obtaining real
poisoned production data, though we make no claim about how universal it
is.

Section~\ref{sec:results} reported a measured failure: at $10\%$
poisoning on CIFAR-10, clustering TPR is $0.000$ and spectral AUROC is
$0.545$, both directly observed, not modeled. The guard-threshold
explanation offered for \emph{clustering's} failure --- the
minority-cluster guard $\tau$ suppressing a class at $50\%$ measured
contamination --- is consistent with the confirmed contamination and
flag-count measurements, which confirm that zero samples were flagged. What remains unconfirmed, as discussed in
Section~\ref{sec:results}, is whether $k$-means's underlying two-way
split, had the guard allowed it through, would have separated poisoned
from clean samples; PneumoniaMNIST's near-total overlap between flagged
and truly poisoned samples is evidence for this in the case where the
guard permits a flag, but no equivalent evidence exists for CIFAR-10 at
$10\%$, where nothing is flagged at all. Critically, this mechanism is
specific to clustering and does not explain spectral signature
analysis's independent degradation at the same configuration: spectral
analysis does not use $\tau$ or any minority-cluster concept at all, so
whatever causes its AUROC to collapse to near-chance here is a separate,
currently unexplained phenomenon, not a downstream consequence of the
clustering guard. We do not have a mechanism to offer for it. Within the
two datasets and configurations tested here, the same fixed $\tau = 0.35$
succeeded on PneumoniaMNIST (target-class contamination $28.0\%$, under
$\tau$) and failed on CIFAR-10 (contamination $50.0\%$, over $\tau$);
whether a fixed threshold can be chosen that handles both regimes well
is not established one way or the other by this study, and is not a
claim we make about datasets beyond the two tested. Making $\tau$
adaptive to the observed class size, or replacing the hard
minority-fraction cutoff with a calibrated statistical test, is
motivated future work for the clustering detector specifically,
contingent on the cluster-alignment question above, and does not address
spectral's separate degradation.

Scope is also limited. The method addresses image-classification
pipelines and the training-time poisoning and backdoor threat class. It
does not address inference-time adversarial examples, model inversion,
prompt injection against multimodal systems, or model drift
exploitation, each named as a distinct concern in sector guidance.
Whether the framework-referenced output is usable in practice by
healthcare security teams is a design intent, not a finding: no user
study has been conducted.

Extending evaluation toward adaptive, adversarially aware poisoning, and
toward pilot deployment with healthcare or critical-infrastructure
partners, is future work.

\paragraph{Scope of what this study verified.} The contamination
fractions, the CIFAR-10 10\% flag counts, and PneumoniaMNIST's exact
class balance reported in Section~\ref{sec:results} were confirmed
directly from each run's saved output, not estimated or assumed; the
seed-dependent clean-data false-positive pattern on CIFAR-10 discussed
there was found in the course of that verification and is not otherwise
explained. Two further limits are worth stating plainly rather than
leaving implicit. First, agreement's precision/recall trade-off against
clustering alone (Section~\ref{sec:results}) is characterized by the
single run reproduced in Figure~\ref{fig:report}, not by aggregating
recall across all 40 runs; the reported conclusion --- a recall cost
with no precision gain in that instance --- should be read as
illustrative of the trade-off the agreement rule can incur, not as a
comprehensive evaluation of it across poisoning rates and datasets.
Second, the assurance report reproduced verbatim in Figure~\ref{fig:report}
is preserved exactly as \texttt{aegis-scan} generated it, including an
internal, non-standard display name for technique AML.T0018
(``Manipulate AI Model''\,) that does not match the \emph{Backdoor ML
Model} terminology this paper uses throughout (Section~\ref{sec:impl}).
The technique ID is correct and unaffected; the reproduced figure is
historical tool output from before this discrepancy was identified, and
is left unedited on that basis rather than corrected after the fact.
The underlying display-string bug in the tool's source is otherwise
unresolved by this paper.

\section{Conclusion}

This paper specifies a score-level fusion of spectral signature analysis
and activation clustering for backdoor detection, releases an
open-source implementation, and evaluates it against each component
detector alone on a medical imaging benchmark and a non-medical
benchmark at four poisoning rates over five seeds, with findings
expressed in NIST AI RMF and MITRE ATLAS terms. Fusion improves on
spectral signature analysis alone on the medical imaging benchmark at
every attacked poisoning rate tested; superiority over clustering alone
was not established (Section~\ref{sec:results}). On the non-medical benchmark at the highest
tested poisoning rate, both component detectors independently fail ---
clustering's true-positive rate and spectral's AUROC both collapse ---
so fusion collapses with them, despite the underlying attack succeeding
almost perfectly. We offer an explanation for clustering's failure,
consistent with but not confirmed by the evidence gathered so far;
spectral's independent failure at the same configuration remains
unexplained. A security-assurance tool's blind spots are as important to
its evidentiary value as its successes, which is why this failure mode
is reported as a primary finding here rather than omitted.

\section*{Disclosure of tool use}

Generative AI language tools were used to assist with drafting,
literature search, and editing of this manuscript, and with building the
evaluation tooling described in Section~\ref{sec:impl}. All experiments
were executed by the author on the author's own machine; the author
verified the reported results against the tool's raw output files and
takes full responsibility for the contents of this work, irrespective of
how any part of it was generated.

\section*{Data and code availability}

The implementation is available at
\url{https://github.com/tsurace37/aegis-scan}
under the Apache License 2.0. All benchmarks used are public.

\end{document}